\documentstyle[multicol,twoside,eqsecnum,epsf,aps,prb]{revtex}

\newcommand{\ms}{\noalign{\vspace{3pt plus2pt minus1pt}}}
\newcommand{\hs}{\noalign{\vspace{12pt plus2pt minus2pt}}}

\begin{document}
\draft
\title{Transverse-field Ising spin chain with inhomogeneous disorder}

\author{Dragi Karevski}
\address{Laboratoire de Physique des Mat\'eriaux, Universit\'e Henri
Poincar\'e (Nancy 1), B.P. 239,\\ F-54506 Vand\oe uvre l\`es Nancy cedex,
France}
\author{R\'obert Juh\'asz}
\address{Institute for Theoretical Physics, Szeged University, H-6720 Szeged,
Hungary\\ and Research Institute for Solid State Physics and Optics, P.O.
Box 49, H-1525 Budapest, Hungary}
\author{Lo\"{\i}c Turban}
\address{Laboratoire de Physique des Mat\'eriaux, Universit\'e Henri
Poincar\'e (Nancy 1), B.P. 239,\\ F-54506 Vand\oe uvre l\`es Nancy cedex,
France}
\author{Ferenc Igl\'oi\cite{perm}}
\address{Research Institute for Solid State Physics and Optics, P.O.
Box 49, H-1525 Budapest, Hungary\\ and Laboratoire de Physique des
Mat\'eriaux, Universit\'e Henri
Poincar\'e (Nancy 1), B.P. 239,\\ F-54506 Vand\oe uvre l\`es Nancy cedex,
France}

\date{February 9, 1999}

\maketitle

\begin{abstract}
We consider the critical and off-critical properties at the boundary of
the random transverse-field Ising spin chain when the distribution
of the couplings and/or transverse fields, at a distance $l$ from the
surface, deviates from its uniform bulk value by terms of order
$l^{-\kappa}$ with an amplitude $A$. Exact results are obtained using a
correspondence between the surface magnetization of the model and the
surviving probability of a random walk with time-dependent absorbing
boundary conditions. For slow enough decay, $\kappa<1/2$, the
inhomogeneity is relevant: Either the surface stays ordered at the bulk
critical point or the average surface magnetization displays an essential
singularity, depending on the sign of $A$. In the marginal situation,
$\kappa=1/2$, the average surface magnetization decays as a power law
with a continuously varying, $A$-dependent, critical exponent which is
obtained analytically. The behavior of the critical and off-critical
autocorrelation functions as well as the scaling form of the probability
distributions for the surface magnetization and the first gaps are
determined through a phenomenological scaling theory. In the  Griffiths
phase, the properties of the Griffiths-McCoy singularities are not
affected by the inhomogeneity. The various results are checked using
numerical methods based on a mapping to free fermions.
\end{abstract}


\begin{multicols}{2}
\narrowtext

\section{Introduction and summary}

Quantum systems with quenched disorder have
received much attention recently, due to their unusual static and
dynamical properties.\cite{rieger97a}  Many of these features 
can be observed in one-dimensional models for which several exact and 
conjectured results
are available in the case of homogeneous disorder. Among these models we
shall consider the random transverse-field Ising model (RTIM) defined by
the Hamiltonian
\begin{equation}
{\cal H}=-\frac{1}{2}\sum_l[J_l\sigma_l^x\sigma_{l+1}^x+h_l
\sigma_l^z]\,,
\label{e1.1}
\end{equation}
where $\sigma_l^x$ and $\sigma_l^z$ are Pauli matrices at site $l$.
The exchange couplings $J_l$ and the transverse fields $h_l$ are
random variables with probability distributions $\pi_l(J)$ and
$\rho_l(h)$, respectively, which are independent from each other but
may depend on the position $l$. Note that in one dimension the
couplings and the fields can be taken positive without loss of
generality.

The homogeneously disordered model, with position-independent
distributions $\pi_l(J)=\pi(J)$ and $\rho_l(h)=\rho(h)$, has received
much attention, especially after Fisher has obtained new
striking results about static critical properties and equal-time
correlations, using a renormalization group method based on a decimation
procedure.\cite{fisher92} Many of Fisher's results have been later verified by
numerical methods and some other analytical, scaling and numerical results
have been obtained concerning the dynamical properties of the RTIM and the
behavior of different probability 
distributions.\cite{igloi97a,rieger97b,igloi98a,igloi98b,igloi99a,young96}

For the homogeneous model in (\ref{e1.1}) one defines the
quantum control parameter as
\begin{equation}
\delta=[\ln h]_{\rm av}-[\ln J]_{\rm av}\,,
\label{e1.2}
\end{equation}
where $[\dots]_{\rm av}$ denotes an average over quenched disorder. The
system is in the ferromagnetic (paramagnetic)
phase when the couplings are in average stronger (weaker) than the
transverse fields, thus for $\delta<0$ ($\delta>0$). The critical point
corresponds to $\delta=0$.

In this paper we concentrate on the surface properties of the RTIM. The
average surface magnetization, $[m_s(\delta)]_{\rm av}$, which can be
studied using a mapping to a problem of surviving random walks 
(RW),\cite{igloi98a} behaves as $[m_s(\delta)]_{\rm av}\sim-\delta$ 
close  to the critical point, whereas the correlation length diverges 
as $[\xi]_{\rm av}\sim\vert\delta\vert^2$. Thus the corresponding critical 
exponents for the average quantities are:
\begin{equation}
\beta_s=1,\ \ \ \nu=2\,.
\label{e1.3}
\end{equation}
The scaling is strongly anisotropic at the critical point of the
homogeneous RTIM: The time scale set by the relaxation time $\tau_r$ and
the length scale $\xi$ are related through
\begin{equation}
\ln\tau_r\sim\xi^{1/2}\,.
\label{e1.4}
\end{equation}
As a consequence, the imaginary-time spin-spin autocorrelations are
logarithmically slow at the critical point with
\begin{equation}
\left[ G_s(\tau)\right]_{\rm av}\sim(\ln\tau)^{-1}\,,\ \ \ \delta=0\,,
\label{e1.5}
\end{equation}
for the surface spins. In the Griffiths phase, on the
paramagnetic side of the critical point, the
autocorrelations are still anomalous and decay like a power,
\begin{equation}
[G(\tau)]_{\rm av}\sim\tau^{-1/z(\delta)}\,,\ \ \ 0<\delta<\delta_G\,,
\label{e1.6}
\end{equation}
with a dynamical exponent $z(\delta)$ given by the positive root of
the following equation:\cite{igloi98b}
\begin{equation}
\left[\left({J \over h}\right)^{1/z}\right]_{\rm av}=1\,.
\label{e1.7}
\end{equation}
As shown recently,\cite{igloi99a} all the singular quantities in the Griffiths phase
(susceptibility, non-linear susceptibility, energy-density
autocorrelations, etc) involve the dynamical exponent $z(\delta)$.

In many physical systems the disorder is not homogeneous. For example a
free surface or an internal defect line may locally
modify the distribution of the couplings and/or fields. Here we consider
surface induced inhomogeneities which are characterized by a
power-law variation in the probability distribution: $\pi_l(J)-\pi(J)
\sim l^{-\kappa}$ and/or $\rho_l(h)-\rho(h) \sim l^{-\kappa}$, for $l \gg
1$, such that the local control parameter in (\ref{e1.2}) varies as
\begin{equation}
\delta(l)=\delta-Al^{-\kappa}\,.
\label{e1.8}
\end{equation}
Our choice for the functional form of the inhomogeneous disorder is
analogous to the variation of the couplings in the so-called
extended surface defect problem first introduced by Hilhorst and 
van Leeuwen (HvL) for the two-dimensional classical Ising
model.\cite{hilhorst81}  This type of inhomogeneity has been later studied for
other models and different geometries. For a review see 
Ref.\onlinecite{igloi93}.

Informations about the critical properties of the RTIM with
inhomogeneous surface disorder have been obtained by combining different
approaches: The calculation of the surface magnetization can be mapped
onto a problem of surviving RW's, allowing us to deduce some exact results.
The asymptotic behavior of the surface autocorrelation function and the
form of the probability distribution for different quantities follow
from a phenomenological scaling theory. They have been checked through large
scale numerical calculations using the free fermion formulation of the RTIM.

\end{multicols}
\widetext
\begin{table}
\squeezetable
\caption{Summary of the surface critical properties of the inhomogeneous
RTIM.}
\begin{tabular}{ccccccc}
&$\ln\tau_r$&$P_\epsilon(\ln \epsilon_1,L)$&$[m_s(\delta)]_{\rm av}$
&$P_m(\ln m_s,L)$&$\lim_{v\to0}\widetilde{P}_m(v)$&$[G_s(\tau)]_{\rm av}$\\
\ms
\tableline
\ms
$\kappa >1/2$&$\sim\xi^{1/2}$
&${1\over L^{1/2}}\widetilde{P}_\epsilon\left[{\ln\epsilon_1\over
L^{1/2}}\right]$&$\sim\vert\delta\vert$
&${1\over L^{1/2}}\widetilde{P}_m\left[{\ln m_s\over L^{1/2}}\right]$
&$\sim$const&$\sim(\ln\tau)^{-1}$\\
\hs
$\kappa=1/2$&$\sim\xi^{1/2}$
&${1\over L^{1/2}}\widetilde{P}_\epsilon\left[{\ln\epsilon_1\over
L^{1/2}}\right]$&$\sim\vert\delta\vert^{\beta_s(A)}$
&${1\over L^{1/2}}\widetilde{P}_m\left[{\ln m_s\over L^{1/2}}\right]$
&$\sim v^{1-\beta_s(A)}$&$\sim(\ln\tau)^{-\beta_s(A)}$\\
\hs
$\kappa<1/2$&$\sim\xi^{1-\kappa}$
&${1\over L^{1-\kappa}}\widetilde{P}_\epsilon\left[{\ln\epsilon_1\over
L^{1-\kappa}}\right]$&---
&${1\over L^{1-\kappa}}\widetilde{P}_m\left[{\ln m_s\over
L^{1-\kappa}}\right]$
&---&---\\
\ms
$A>0$&---&---
&$\sim\vert A\vert^{1\over1-2\kappa},\,\delta=0$&---&$v^{-1}$&$\sim$ const\\
\ms
$A<0$&---&---&$\sim\exp[-{\rm const}\,
\vert\delta\vert^{-(1-2\kappa)/\kappa}]$ &---&0&$\sim\exp[-{\rm
const}\,(\ln\tau)^{1-2\kappa\over1-\kappa}]$\\ \ms
 \end{tabular}
 \label{table.1}
 \end{table}
\begin{multicols}{2}
\narrowtext

Our main results are summarized in Table I. The surface critical behavior of
the inhomogeneous RTIM depends on the value of the decay exponent $\kappa$.
For fast enough decay of the inhomogeneity, i.e., for $\kappa>1/2$,
the surface critical properties of the model are the same as for the
homogeneous RTIM. Specifically, the critical exponents keep the values
given in Eq.~(\ref{e1.3}) and relations~(\ref{e1.4}) and~(\ref{e1.5})
remain valid.

In the borderline case, $\kappa=1/2$, the effect of the inhomogeneity is
marginal: the average surface magnetization has a
distribution-dependent critical exponent, $\beta_s(A)$, which has been
calculated analytically. This exponent governs also the logarithmic decay
of the surface spin-spin autocorrelation function, like in Eq.~(\ref{e1.5})
for the homogeneous case, but with a different value. The scaling relation
between $\tau_r$ and $\xi$ keeps the form given in Eq.~(\ref{e1.4}) for the
homogeneous model. The same is true of the scaling behavior of the
probability distributions for the energy gap $\epsilon$, and the
surface magnetization $m_s$, at large system size $L$. The
scaling function of the latter, however, has a different asymptotic behavior,
which is connected to the different $\delta$-dependence of the average
surface magnetization in the two cases.

For a slower decay, such that $\kappa<1/2$, the scaling relation in
Eq.~(\ref{e1.4}) is modified and leads to a new scaling form for the
probability distributions of the energy gap and the suface magnetization. For
enhanced surface couplings, $A>0$, the average surface magnetization remains
finite at the bulk critical point. It vanishes as a power of $A$ when $A$ goes
to zero from above. For weakened surface couplings, the average surface
magnetization displays an essential singularity in $\delta$ instead of a
power law, whereas the surface autocorrelation function has an enhanced
power-law decay.

In the numerical calculations we used two types of distributions for the
disorder. In the inhomogeneous binary distribution, the couplings take the
values $\Lambda>1$ and $\Lambda^{-1}$ with probability
$p_l=\frac{1}{2}(1+A_bl^{-\kappa})$ and $q_l=1-p_l$, respectively, while the
transverse field remains constant:
\begin{eqnarray}
\pi_l(J)&=&p_l\delta(J-\Lambda)+q_l\delta(J-\Lambda^{-1})\,,
\nonumber\\
\rho(h)&=&\delta(h-h_0)\,.
\label{e1.9}
\end{eqnarray}
According to Eq.~(\ref{e1.2}) $h_0=1$ at the bulk critical point and the local
control parameter in Eq.~(\ref{e1.8}) involves the parameter
$A=A_b\ln\Lambda$. In the bulk Griffiths phase, $1<h_0<\Lambda$, the
dynamical exponent, which follows from Eq.~(\ref{e1.7}), is the solution of
the implicit equation:
\begin{equation}
h_0^{1/z}=\cosh\left(\ln\Lambda\over z\right)\,.
\label{e1.10}
\end{equation}

In the uniform distribution, both the couplings and the fields have
rectangular distributions:
\begin{eqnarray}
\pi_l(J)&=&\pi(J)=\left\{\begin{array}{ll}1&\ \ \mbox{if $0<J<1$}\\
                0&\mbox{otherwise}\end{array}\right.\,,\nonumber\\
\rho_l(h)&=&\left\{\begin{array}{ll}[h_0(l)]^{-1}&\ \ \mbox{if $0<h<h_0(l)$}\\
                0&\mbox{otherwise}\end{array}\right.\,,
\label{e1.11}
\end{eqnarray}
where the inhomogeneity now affects the distribution of the fields with
$h_0(l)=h_0-A_ul^{-\kappa}$. The bulk critical point is still at $h_0=1$ 
whereas $A=A_u$ in the expression of the local control parameter 
in Eq.~(\ref{e1.8}). The dynamical exponent follows from
\begin{equation}
z\ln\left(1-z^{-2}\right)=-\ln h_0\,,
\label{e1.12}
\end{equation}
and the domain of the Griffiths phase now extends to $1<h_0<\infty$.

The structure of the paper is the following: In Sec.~II we use the free
fermion formulation of the RTIM to express different surface
quantities. The relation between the surface magnetization of
the inhomogeneous RTIM and an absorbing RW problem is treated in Sec.~III.
It is exploited there to develop a relevance-irrelevance criterion. 
Our results are presented in Sec.~IV  and discussed in Sec.~V.

\section{Free-fermionic expression of surface quantities}

We start by considering the imaginary-time surface autocorrelation function
of the RTIM
\begin{eqnarray}
G_s(\tau)&=&\langle0\vert\sigma_1^x(\tau)\sigma_1^x(0)
\vert0\rangle\nonumber\\
&=&\sum_i\vert\langle i\vert\sigma_1^x\vert0\rangle\vert^2
\exp\left[-\tau(E_i-E_0)\right]\,,
\label{e2.1}
\end{eqnarray}
where $\vert0\rangle$ and $\vert i\rangle$ denote the ground state and the
$i$-th excited state of ${\cal H}$ in Eq.~(\ref{e1.1}) with energies $E_0$
and $E_i$, respectively. With symmetry-breaking boundary conditions, i.e.,
with a fixed spin at the right end of the chain, $\sigma_L^x=\pm1$, the
ground state of the system is degenerate and the autocorrelation function
asymptotically behaves as $\lim_{\tau\to\infty}G_s(\tau)=m_s^2$, so that the
surface magnetization is given by:
\begin{equation}
m_s=\langle1\vert\sigma_1^x\vert0\rangle\,.
\label{e2.2}
\end{equation}

In order to calculate $m_s$ and other quantities we use the method of Lieb,
Schultz and Mattis\cite{lieb61} and transform ${\cal H}$ into
a free fermion Hamiltonian,
\begin{equation}
{\cal H}=\sum_{q=1}^L\epsilon_q\left(\eta_q^\dagger\eta_q
-\frac{1}{2}\right)\,,
\label{e2.3}
\end{equation}
where the $\eta_q^\dagger$'s ($\eta_q$'s) are fermion creation (annihilation)
operators. For free boundary conditions the fermion excitation
energies $\epsilon_q$ are obtained through the diagonalization of the
following $2L\times2L$ tridiagonal matrix:\cite{igloi96}
\begin{equation}
T= \left(
\matrix{
 0  & h_1 &     &       &       &       &     \cr
h_1 &  0  & J_1 &       &       &       &     \cr
 0  & J_1 &  0  & h_2   &       &       &     \cr
    &     & h_2 &  0    &\ddots &       &     \cr
    &     &      &\ddots&\ddots &J_{L-1}&     \cr
    &     &      &      &J_{L-1}&   0   & h_L \cr
    &     &      &      &       &  h_L  &  0  \cr
}
\right)\,.
\label{e2.4}
\end{equation}
The components of the eigenvectors ${\bf V}_q$ are written as
$V_q(2l-1)=-\phi_q(l)$ and $V_q(2l)=\psi_q(l)$, $l=1,2,\dots,L$. Changing
$\bbox{\phi}_q$ into $-\bbox{\phi}_q$ one obtains the eigenvector
corresponding to $-\epsilon_q$, thus we confine ourselves to
that part of the spectrum corresponding to $\epsilon_q\ge0$
which contains all the needed information.\cite{igloi96}  Fixing the surface
spin at $l=L$ amounts to take $h_L=0$. Then  $\epsilon_1=0$ and the ground
state of the Hamiltonian is degenerate.

The surface magnetization and the spin-spin autocorrelation function in the
free-fermion description are calculated using Wick's theorem and can be
expressed in terms of the first component of the eigenvector corresponding
to the first excitation as\cite{peschel84}
\begin{equation}
m_s(L)=\phi_1(1)=\left[1+\sum_{l=1}^{L-1}\prod_{j=1}^l\left({h_j\over
J_j}\right)^2\right]^{-1/2}\! ,
\label{e2.5}
\end{equation}
and
\begin{equation}
G_s(\tau)=\sum_q\vert\phi_q(1)\vert^2\exp(-\tau\epsilon_q)\,.
\label{e2.6}
\end{equation}
The first gap of the model for a free chain, $\epsilon_1(L)$, is related to
the value of the surface magnetization (\ref{e2.5}). It can be shown that,
provided it vanishes faster than $L^{-1}$, the first gap $\epsilon_1(L)$
satisfies the asymptotic relation:\cite{igloi97b,igloi98a}
\begin{equation}
\epsilon_1(L)\sim m_s(L){\overline m}_s(L)h_L\prod_{l=1}^{L-1}{h_l\over
J_l}\,.
\label{e2.7}
\end{equation}
Here $m_s(L)$ and ${\overline m}_s(L)$ denote the finite-size surface
magnetizations at both ends of the chain and ${\overline m}_s(L)$ follows from
the substitution $h_j/J_j\leftrightarrow h_{L-j}/J_{L-j}$ in
Eq.~(\ref{e2.5}). One may notice that Eqs.~(\ref{e2.5}) and~(\ref{e2.7}) can
be rewritten into the equivalent forms
\begin{eqnarray}
m_s(L)&=&\overline{m}^d_s(L)\prod_{l=1}^{L-1}{J_l\over h_l}\nonumber\\
\epsilon_1(L)&\sim & m_s(L)m^d_s(L)h_L\,,\ \ \ m_s(L)\ge m^d_s(L)\,,
\label{e2.8}
\end{eqnarray}
where $m^d_s(L)$ and $\overline{m}^d_s(L)$ are the finite-size surface
magnetizations at both ends with dual interactions, i.e., when the fields and
couplings are interchanged, $h_j \leftrightarrow J_j$. The condition for the
validity of relation (\ref{e2.7}), $\epsilon_1(L)\le L^{-1}$, is verified
when the local couplings are in average not weaker than the fields, from
which the condition $m_s(L)\ge m^d_s(L)$ in Eq.~(\ref{e2.8}) follows.

\section{Random-walk description of the surface magnetization}

The surface magnetization, which is related in Eq.~(\ref{e2.5}) to a sum of
products of the ratios $(h_j/J_j)^2$, can be
evaluated by exploiting a random walk analogy, developed in
Ref.\onlinecite{igloi98a}. We shall present the method by first considering
the homogeneous random model at the critical point $h_0=1$ with the
extreme binary distribution, i.e., in the limit $\Lambda\to\infty$ with
$p_l=q_l=1/2$ in Eq.~(\ref{e1.9}). With this distribution, the surface
magnetization of a sample vanishes whenever a product $\prod_{j=1}^l
J_j^{-2}$ ($l=1,2,\dots,L-1$) is infinite, i.e., when there are more
$\Lambda^{-1}$ than $\Lambda$ couplings on any of the
intervals $[1,l]$. Otherwise, when this condition is not
fullfilled, i.e., when for any of the intervals $[1,l]$ the number of
$\Lambda$ couplings is not smaller than the number of $\Lambda^{-1}$, the
surface magnetization of the sample is nonvanishing, $m_s(L)=O(1)$.

\begin{figure}[t]  
\epsfxsize=7.6cm
\begin{center}
\hspace*{-13truemm}\mbox{\epsfbox{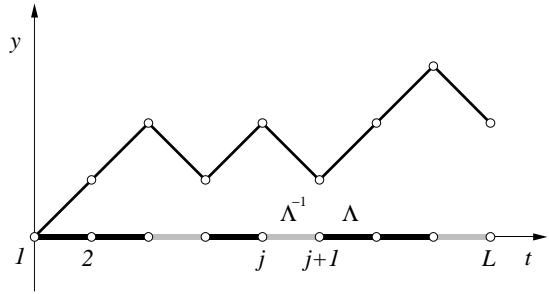}}
\end{center}
\caption{Random walk correspondance of the distribution of couplings.}
\label{fig-hvl-rand-rw}
\end{figure}

The distribution of the couplings in a sample can be put in correspondence
with a RW, $y(t)$, the time axis $t$ corresponding to the position
$j$ along the spin chain. As shown in Fig.~\ref{fig-hvl-rand-rw}, the walker
starts at $y=0$, its $j$-th step is in the positive (negative) $y$-direction
when $J_j=\Lambda$ ($\Lambda^{-1}$) and absorbing boundary
conditions are assumed for $y<0$. As a consequence, the surface magnetization
of a sample is nonvanishing only when the corresponding walk survives until
$t=L$, i.e., if it never crosses the time axis during $L-1$ steps. Thus the
fraction of $L$-step surviving walks, $\left.P_{\rm surv}(t)\right|_{t=L}\sim
L^{-1/2}$, gives the fraction of {\it rare events} for which a sample has a
nonvanishing surface magnetization. Since the average value of the surface
magnetization  is determined by the rare events, we have the correspondence:
$[m_s(L)]_{\rm{av}}\sim\left. P_{\rm surv}(t)\right|_{t=L}\sim L^{-1/2}$ from
which the value of the critical exponent $x_s=\beta_s/\nu=1/2$ follows, in
agreement with Eq.~(\ref{e1.3}).

Let us consider the average of the logarithm of the products in
Eq.~(\ref{e2.5}) which may be written as
\begin{equation}
\left[\ln\prod_{j=1}^l\left({h_j\over J_j}\right)^2\right]_{\rm av}
=2\sum_{j=1}^l\left([\ln h_j]_{\rm av}-[\ln J_j]_{\rm av}\right)
=2\delta\, l\,,
\label{e3.1}
\end{equation}
using the definition of the control parameter in Eq.~(\ref{e1.2}).
In the RW picture this quantity is proportional to the average position of
the walker after $l$ steps for a free walk and vanishes at criticality.
In the off-critical situation, $\delta\ne0$, the average position of the
walker grows linearly with $l$, which is equivalent
to having a nonzero bias $\delta_w=q_w-p_w\ne0$ where $p_w$ ($q_w$)
denotes the probability that the walker makes a step in the positive
(negative) $y$-direction. Thus the correspondence
between RTIM and RW still applies in the off-critical situation with:
\begin{equation}
[m_s(\delta,L)]_{\rm av}\sim
\left.P_{\rm surv}(\delta_w,t)\right|_{t=L}\,,\ \ \ \delta\sim\delta_w\,.
\label{e3.2} \end{equation}
{}From the known behavior of the surviving probability, one deduces the
values of the RTIM critical exponents given in Eq.~(\ref{e1.3}).

For the inhomogeneous RTIM the local quantum control parameter has a smooth
position dependence which, at the bulk critical point, is given by
$\delta(l)=-Al^{-\kappa}$ according to Eq.~(\ref{e1.8}). The corresponding
RW has a locally varying bias with the same type of asymptotic dependence,
$\delta_w(l)=-A_wl^{-\kappa}$. Consequently the average motion of the
walker is parabolic:
\begin{equation}
y_p(t)=-\sum_{l=1}^t\delta_w(l)={A_w\over1-\kappa}t^{1-\kappa}\,.
\label{e3.3}
\end{equation}
Under the change of variable $y(t)\to y(t)-y_p(t)$, the surviving
probability of the inhomogeneously drifted walker is also the
surviving probability of an unbiased walker, however with a time-dependent
absorbing boundary condition at $y(t)<-y_p(t)$.

The surviving probability of a random walker with
time-dependent absorbing boundaries has already been studied in the
mathematical\cite{uchiyama80} and physical\cite{turban92,kaprivsky96} literature. In a continuum 
approximation, it
follows from the solution of the diffusion equation with appropriate
boundary conditions,
\begin{equation}
{\partial\over\partial t}P(y,t)=D{\partial^2\over\partial y^2}P(y,t)\,,
\ \ P[-y_p(t),t]=0\,.
\label{e3.4}
\end{equation}
Here $P(y,t)$ is the probability density for the position of the walker at
time $t$ so that the surviving probability is given by
\begin{equation}
P_{\rm surv}(t)=\int_{-y_p(t)}^{\infty}\!\! P(y,t)dy\,.
\label{e3.5}
\end{equation}
The behavior of the surviving probability depends on the value of
the decay exponent $\kappa$. For $\kappa>1/2$, the drift of the absorbing boundary
in Eq.~(\ref{e3.3}) is slower then the diffusive motion of the
walker, typically given by
\begin{equation}
y_d(t)\sim (D t)^{1/2}\,,
\label{e3.6}
\end{equation}
thus the surviving probability behaves as in the static case. When
$\kappa<1/2$, the drift of the absorbing boundary is faster than the
diffusive motion of the walker and leads to a new behavior for the surviving
probability. For $A_w>0$, since the distance to the moving boundary grows in
time, the surviving probability approaches a finite limit. On the contrary,
for $A_w<0$, the boundary moves towards the walker and the surviving
probability decreases with a fast, stretched-exponential
dependence on $t$. Finally, in the borderline case $\kappa=1/2$
where the drift of the boundary and the diffusive motion have the same
dependence on $t$, like in the static case the surviving probability decays as
a power, $P_{\rm surv}(\delta_w,t)\sim t^{-\theta(A_w)}$, however with a
continuously varying critical exponent.

Now turning back to the problem of the inhomogeneous RTIM, the RW analogy
can be used to formulate a relevance-irrelevance criterion: According to
the preceding discussion, the inhomogenous surface
disorder is a relevant (irrelevant) perturbation for the surface
critical properties of the RTIM when $\kappa<1/2$ ($>1/2$) and a marginal
one when $\kappa=1/2$.

\section{Results}

In this Section we study the surface critical properties of the inhomogeneous
RTIM, considering the size-dependence at criticality as well as the
$\delta$-dependence for various quantities. We treat successively relevant and
marginal perturbations and close with a discussion of the behavior in the Griffiths phase.

In the numerical calculations we used up to $5\;10^{4}$ realizations of disorder
with chain sizes up to $L=2^7$ except for the surface magnetization for which we used
up to $10^6$ realizations with system sizes up to $L=2^{14}$.

\subsection{Critical behavior with relevant inhomogeneity}

\begin{figure}[t] 
\epsfxsize=7.6cm
\begin{center}
\hspace*{-13truemm}\mbox{\epsfbox{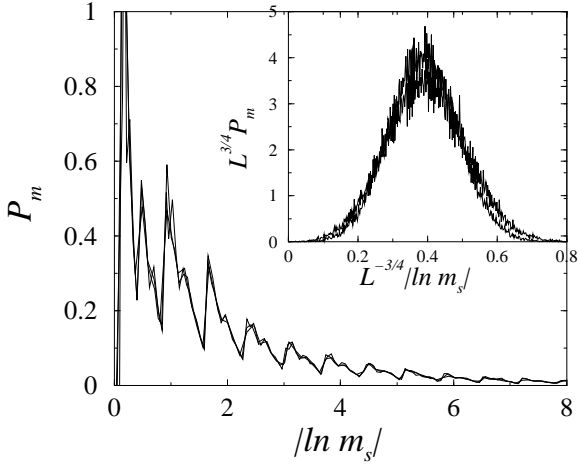}}
\end{center}
\caption{Probability distribution of $\ln m_{s}$ obtained for the binary distribution
with $\kappa=1/4$ for $A=0.4$ and scaling plot for $A=-0.4$ (inset) for chain
sizes $L=2^{10}, 2^{11},2^{12}$.}
\label{fig-hvl-rand-plnms-rel}
\end{figure}

Let us first consider the probability distribution of $\ln m_s$ on finite
samples with length $L$ at the bulk critical point $\delta=0$. According to
Eq.~(\ref{e2.8}), $[\ln m_s]_{\rm av}$ is expected to scale as
$[\ln\prod_{l=1}^L(J_l/h_l)]_{\rm av}$ when the perturbation tends to reduce
the surface order, i.e., when $A<0$ in Eq.~(\ref{e1.8}). Thus one obtains:
\begin{eqnarray}
[\ln m_s]_{\rm av}&\sim&\sum_{l=1}^L([\ln J_l]
_{\rm av}-[\ln h_l]_{\rm av}\nonumber\\
&=&A\sum_{l=1}^Ll^{-\kappa}\sim AL^{1-\kappa}\,.
\label{e4.1}
\end{eqnarray}
The typical magnetization, define through
$[\ln m_s]_{\rm av}=\ln[m_s]_{\rm typ}$, has a stretched-exponential
behavior, $[m_s]_{\rm typ}\sim\exp( {\rm const}\, AL^{1-\kappa})$, when $A<0$.
For the probability distribution of $\ln m_s$, Eq.~(\ref{e4.1}) suggests the
following scaling form:
\begin{equation}
P_m(\ln m_s, L)={1\over L^{1-\kappa}}\tilde{P}_m\left[{\ln m_s\over
L^{1-\kappa}}\right]\,.
\label{e4.2}
\end{equation}
Indeed, as shown in Fig.~\ref{fig-hvl-rand-plnms-rel}, the numerical
results are
in agreement with Eq.~(\ref{e4.2}).

The asymptotic behavior of the scaling function $\tilde{P}_m(v)$ as $v\to0$
depends on the signe of $A$. For enhanced surface couplings, $A>0$, there
is a nonvanishing surface magnetization at the bulk critical point as
$L\to\infty$, thus the powers of $L$ in Eq.~(\ref{e4.2}) must cancel 
as seen in Fig.~\ref{fig-hvl-rand-plnms-rel} 
and $\lim_{v\to0}\tilde{P}_m(v)\sim v^{-1}$ so that $P_m(\ln m_s)\sim1/\ln m_s$.
For reduced surface couplings,
$A<0$, $\lim_{v\to0}\tilde{P}_m(v)=0$, indicating a vanishing
surface magnetization.

Next we calculate the {\it average behavior} of the surface magnetization,
which is determined by the rare events with $m_s=O(1)$. Here we use the RW
description of Section III. Starting with $A\sim A_w>0$,
for $A_w\ll1$ one defines the time scale $t^*$ for which the parabolic and
diffusive lengths in Eqs.~(\ref{e3.3}) and~(\ref{e3.6}) are of the same order,
$y_p(t^*)\sim y_d(t^*)$, such that $t^*\sim A_w^{-2/(1-2\kappa)}$. The
surviving probability can be estimated by noticing that, if the walker is not
absorbed up to $t^*$, it will later survive with a finite probability. Thus
$P_{\rm surv}(A_w>0)\sim(t^*)^{-1/2}\sim A_w^{1/(1-2\kappa)}$ and the average
surface magnetization has the same behavior at the bulk critical point:
\begin{equation}
[m_s]_{\rm av}\sim A^{1/(1-2\kappa)},\ \ \kappa<\frac{1}{2},\ \ 0<A\ll1.
\label{e4.3}
\end{equation}
%
\begin{figure}[t] 
\epsfxsize=7.6cm
\begin{center}
\hspace*{-13truemm}\mbox{\epsfbox{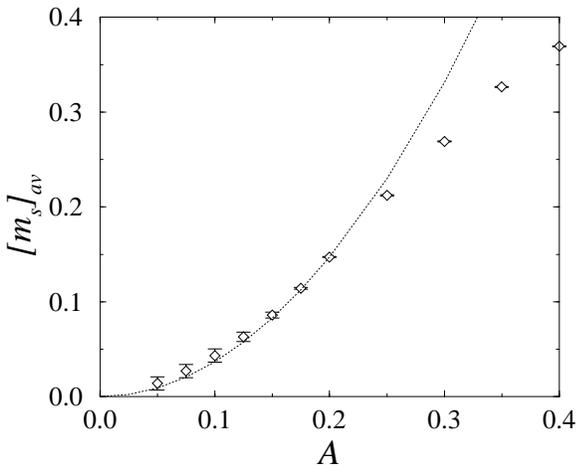}}
\end{center}
\caption{Average surface magnetization as a function of the inhomogeneity
parameter for the binary distribution with $\kappa=1/4$, extrapolated at infinite size. 
The dotted line corresponds to the asymptotic expression in Eq.(\ref{e4.3}).}
\label{fig-hvl-rand-ms-av-fo}
\end{figure}

This result is in agreement with the numerical calculations as shown in
Fig.~\ref{fig-hvl-rand-ms-av-fo}.
The error bars give the precision of the   
extrapolation procedure.

\begin{figure}[t] 
\epsfxsize=7.6cm
\begin{center}
\hspace*{-13truemm}\mbox{\epsfbox{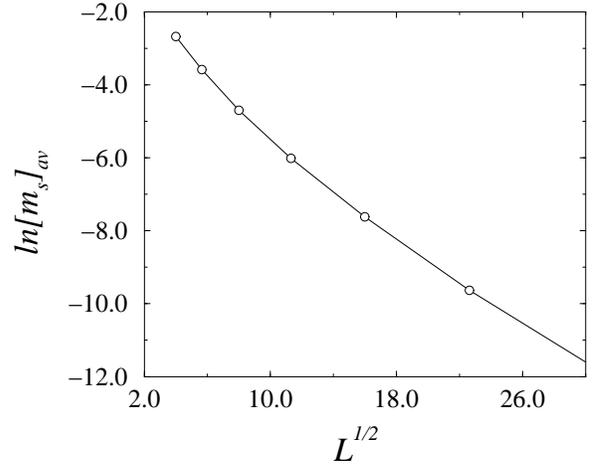}}
\end{center}
\caption{$\ln [m_s]_{\rm{av}}$ as a function of $L^{1-2\kappa}$ for the binary distribution with $\kappa=1/4$
and $A=-0.4$. The asymptotic linear dependence is in agreement with Eq.(\ref{e4.4}).}
\label{fig-hvl-rand-ms-av-rel}
\end{figure}

In the case of a shrinking interval between the walker and the absorbing
wall, $A_w<0$, the leading behavior of the surviving probability can be
estimated by looking for the fraction of walks with
$y(t)>-y_p(t)$ which is given by $P_{\rm surv}(t)\sim\exp[-{\rm const}\,
A_w^2\, t^{1-2\kappa}]$. Thus, for reduced surface couplings, the average
surface magnetization has the following finite-size behavior at the critical
point:
\begin{equation}
[m_s]_{\rm{av}}\sim\exp\left(-{\rm const}\,A^2\,
L^{1-2\kappa}\right),\kappa<\frac{1}{2},A<0\,.
\label{e4.4}
\end{equation}
Numerical results are shown in Fig.~\ref{fig-hvl-rand-ms-av-rel}.
One can notice that even for such large system size as $L=2^9$ the 
linear asymptotic regime is not completely reached.
One may notice the different size dependence of the typical and
average surface magnetizations at criticality in Eqs.~(\ref{e4.2})
and~(\ref{e4.4}), respectively.

To obtain the $\delta$-dependence of the average surface magnetization in
the thermodynamic limit, we first determine the typical size of the surface
region $l_s$ which is affected by the inhomogeneity for $\delta<0$ and $A<0$.
It is given by the condition that quantum fluctuation ($\sim\delta$) and
inhomogeneity ($Al^{-\kappa}$) contributions to the energy are of the same
order, from which the relation $l_s\sim\vert A\vert^{1/\kappa}\vert
\delta\vert^{-1/\kappa}$ follows.
Inserting $L\sim l_s$ in Eq.~(\ref{e4.4}), we obtain
\begin{equation}
[m_s]_{\rm{av}}\sim\exp[-{\rm const}\,\vert A\vert^{1/\kappa}
\,\vert\delta\vert^{-(1-2\kappa)/\kappa}],\kappa<\frac{1}{2},A<0,
\label{e4.5}
\end{equation}
as quoted in Table I. The numerical verification of this result is
difficult since, close to the critical point, one has to use large finite
samples, $L\gg l_s$, for which $[m_s(\delta,A)]_{\rm{av}}$ is quite small.

\begin{figure}[t] 
\epsfxsize=7.6cm
\begin{center}
\hspace*{-13truemm}\mbox{\epsfbox{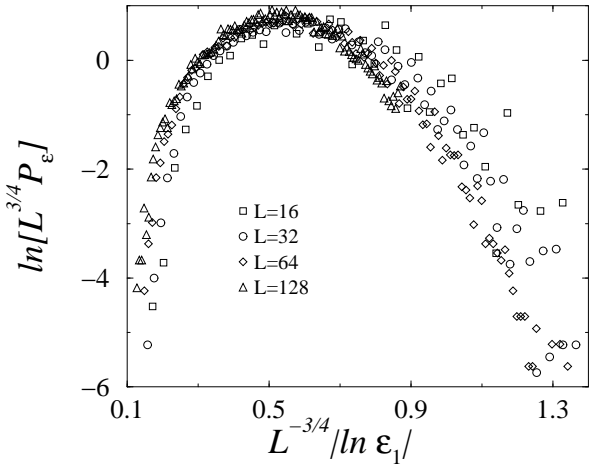}}
\end{center}
\caption{Scaling plot of the probability distribution of the gap for the binary distribution
with $A=0.5$ and $\kappa=1/4$.}
\label{fig-hvl-rand-gap-rel}
\end{figure}

Let us now study the behavior of the energy
gap $\epsilon_1(L)\sim\tau_r^{-1}$.  Here we make use of the relation
(\ref{e2.8}) between the gap and the surface magnetizations, from which the
typical finite-size-scaling behavior $\ln\epsilon_1(L)\sim L^{1-\kappa}$
follows, as for the surface magnetization. Thus the relation between the
time and length scales is
\begin{equation}
\ln\tau_r\sim \xi^{1-\kappa}\,,
\label{e4.6}
\end{equation}
and the probability distribution of the gap is
\begin{equation}
P_\epsilon(\ln\epsilon_1,L)={1\over L^{1-\kappa}}
\tilde{P}_{\epsilon}\left[{\ln\epsilon_1\over L^{1-\kappa}}\right]\,,
\label{e4.7}
\end{equation}
to be compared to the distribution of the surface magnetization in
Eq.~(\ref{e4.2}). Numerical results shown in
Fig.~\ref{fig-hvl-rand-gap-rel} are
in agreement with this scaling result.

\begin{figure}[t]  
\epsfxsize=7.6cm
\begin{center}
\hspace*{-13truemm}\mbox{\epsfbox{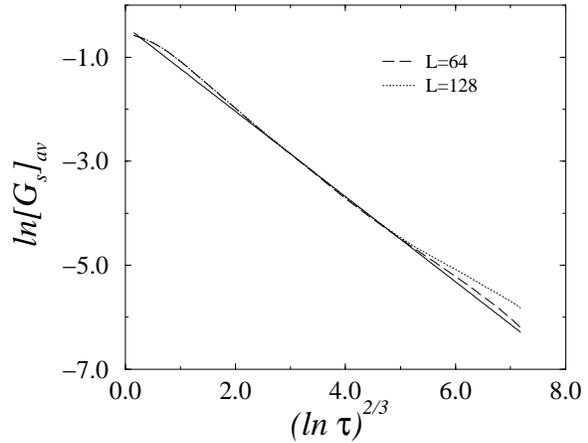}}
\end{center}
\caption{Average surface autocorrelation function for the uniform distribution
with $\kappa=1/4$ and $A=-0.5$. The solid straight line is a guide for the eye.}
\label{fig-hvl-rand-gs-rel}
\end{figure}

In order to obtain an estimate for the average surface autocorrelation
function one may notice that the disorder being strictly correlated along the
time axis, a sample with a finite surface magnetization $m_s=O(1)$ has
also a nonvanishing surface autocorrelation function $G_s(\tau)\sim m_s^2$.
Since the fraction of rare events are the same for the two quantities, the
scaling behavior of $[G_s(L,\tau)]_{\rm{av}}$ can be deduced from the
corresponding relations for the average surface magnetization. For
enhanced surface couplings, $A>0$, according to Eq.~(\ref{e4.3}), $\lim_{\tau
\to\infty}[G_s(\tau)]_{\rm{av}}\sim[m_s(A)]_{\rm{av}}\sim
A^{1/(1-2\kappa)}$ at criticality. For reduced surface couplings, $A<0$, the
finite-size critical behavior follows from Eq.~(\ref{e4.4}) with
$\lim_{\tau\to\infty}[G_s(L,\tau)]_{\rm av}\sim[m_s(L,A)]_{\rm av}\sim
\exp[-{\rm const}\, A^2\, L^{1-2\kappa}]$. Now, using the scaling
relation (\ref{e4.6}), we obtain
\begin{equation}
[G_s(\tau)]_{\rm av}\sim\exp\left[-{\rm const}\,
A^2\,(\ln\tau)^{1-2\kappa\over1-\kappa}\right]\,
\label{e4.8}
\end{equation}
in the thermodynamic limit. This enhanced power-law decay is consistent with
the numerical  results, as shown in Fig.~\ref{fig-hvl-rand-gs-rel}.

\subsection{Critical behavior with marginal inhomogeneity}

In the borderline case, $\kappa=1/2$, several results in the previous
Section have to be modified. We start with the analysis of the average surface
magnetization. For the extreme binary distribution, with $\Lambda\to\infty$ in
Eq.~(\ref{e1.9}), there is a one-to-one correspondence between samples with a
nonvanishing surface magnetization and RW's with surviving character, as
explained in Section III. Thus for this distribution one can deduce $\beta_s$
from the corresponding behavior of the surviving probability in the RW
problem. The differential equation with absorbing boundary conditions in
Eq.~(\ref{e3.4}) can be solved\cite{kaprivsky96} for $\kappa=1/2$ in terms 
of parabolic cylinder functions of order $\nu$, 
$D_{\nu}(x)$.\cite{abramowitz65} For zero bulk bias,
$\delta_w=0$, the surviving probability has the asymptotic dependence
$P_{\rm surv}(t)\sim t^{-\theta}$ and the exponent $\theta$ is such that
\begin{equation}
D_{2\theta}(-2A_w)=0\,,
\label{e4.9}
\end{equation}
taking $D=1/2$ for the diffusion constant in Eq.~(\ref{e3.4}). In the
limiting cases the solution takes the form
\begin{equation}
\theta={1\over2}-\sqrt{{2\over\pi}}\, A_w
\label{e4.10}
\end{equation}
for $|A_w|\ll1$ while for large values of $A_w$ one obtains asymptotically:
\begin{equation}
\theta=\sqrt{{2\over\pi}}\,A_w \exp[-2A_w^2]\,.
\label{e4.11}
\end{equation}
Then the correspondence of Eq.~(\ref{e3.2}) leads to the scaling dimension
of the average surface magnetization,
\begin{equation}
x_m^s={\beta_s\over\nu}=\theta(A_w)\,,
\label{e4.12}
\end{equation}
where the average correlation length exponent $\nu=2$ and $A_w=A_b$ for the extreme binary distribution.

For other inhomogeneous distributions, like in Eqs.~(\ref{e1.9})
and~(\ref{e1.11}), one has to find the relation between the parameter of the
walk $A_w$ and that of the RTIM $A$. The argument goes as follows: At the
bulk critical point, we compare the integrated change in the control parameter
induced by the perturbation at a length scale $l$,
$\Delta(l)=\sum_{j=1}^l[\ln J_j-\ln h_j]_{\rm av}$ to its fluctuation
characterized by $\Gamma^2(l) =\sum_{j=1}^l[(\ln J_j-\ln h_j)^2]_{\rm av}$.
The dimensionless ratio $\Delta(l)/\Gamma(l)$ is equal to $2A_b$ and
$\sqrt{2}A_u$ for the binary and the uniform distribution, respectively.
Since it does not depend on the value of $\Lambda$ for the binary
distribution and since $A_w=A_b$ when $\Lambda\to\infty$, we can identify
this ratio to $2A_w$ which gives:
\begin{equation}
A_w=\left\{\begin{array}{ll}A_b&\ \ \mbox{binary distribution}\\
                {A_u/\sqrt{2}}&\ \ \mbox{uniform distribution}\end{array}
\right.\,.
\label{e4.13}
\end{equation}
%
\begin{figure}[t] 
\epsfxsize=7.6cm
\begin{center}
\hspace*{-13truemm}\mbox{\epsfbox{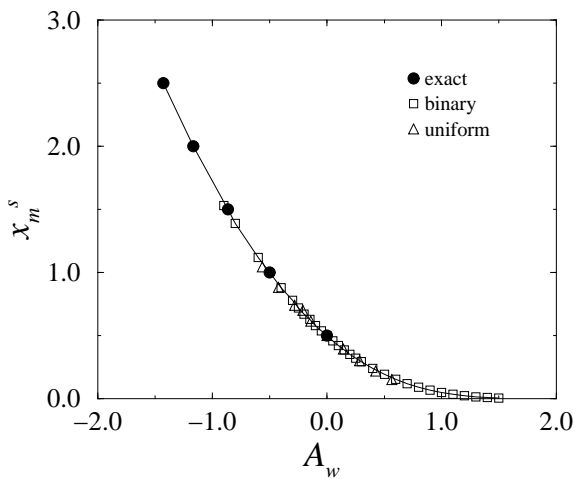}}
\end{center}
\caption{Surface magnetic exponent as a function of the inhomogeneity parameter $A_{w}$.
The line is a guide for the eye.}
\label{fig-hvl-rand-betas-marg}
\end{figure}

The surface magnetization exponent thus depends on the
distribution, however only through the value of the parameter $A_w$ entering
in Eq.~(\ref{e4.12}), the functional form remaining the same.
The average magnetization has been obtained numerically
for the binary and the uniform distribution using $2\;10^{5}$ realizations of disorder for chain 
sizes $L=2^{5}$ to $2^{14}$. The exponents $x_{m}^{s}$ were deduced from an extrapolation of  
two-points approximants using the BST algorithm\cite{henkel88}.
The results are shown in
Fig.~\ref{fig-hvl-rand-betas-marg}.
The exact points in the figure are obtained by solving Eq.~(\ref{e4.9}) for integer and half-integer values of 
$\theta$ where the parabolic cylinder functions are related to Hermite polynomials\cite{abramowitz65}.
The corresponding $A_w$ values are then given by the highest negative zeros of the Hermite polynomial, 
${\rm He}_{2\theta}(-2A_{w})=0$. For $x_{m}^{s}=0$, $A_{w}$ is shifted to $+\infty$ as it can be seen in Eq.~(\ref{e4.11}).
Thus the system does not display surface order at criticality as it is the case for the HvL model for sufficiently enhanced
surface couplings.

\begin{figure}[t]  
\epsfxsize=7.6cm
\begin{center}
\hspace*{-13truemm}\mbox{\epsfbox{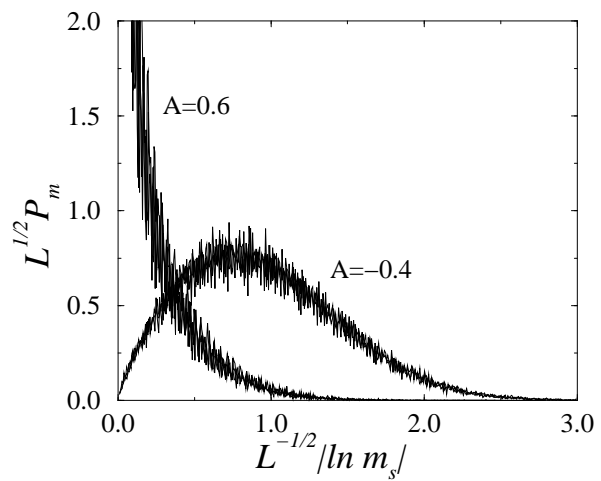}}
\end{center}
\caption{Scaling plot of the probability distribution of $\ln m_{s}$
for the binary distribution with $\kappa=1/2$ for chain 
sizes $L=2^{10},2^{11},2^{12}$.}
\label{fig-hvl-rand-plnms-marg}
\end{figure}

Next we discuss the properties of the probability distribution of $\ln m_s$.
Repeating the argument used above for relevant perturbations, we arrive to the
scaling form of Eq.~(\ref{e4.2}), however with $\kappa=1/2$. We note that
the same scaling form remains valid for irrelevant perturbations, with
$\kappa>1/2$, but the scaling function $\tilde{P}_m(v)$ has different
limiting behaviors when $v\to0$ in the two cases. While for
$\kappa>1/2$ it approaches a constant, in the marginal case it goes to zero as
$\lim_{v\to0}\tilde{P}_m(v)\sim  v^{-2x_m^s(A)+1}$. In this way one
obtains the proper scaling behavior for the average surface magnetization. The
numerical results in Fig.~\ref{fig-hvl-rand-plnms-marg} are in agreement with
this limiting behavior.

For the distribution of the energy gap at the critical point, one can also
repeat the reasoning used for the relevant case and both the distribution
function in Eq.~(\ref{e4.7}) and the scaling relation in Eq.~(\ref{e4.6})
stay valid, with $\kappa=1/2$.

\begin{figure}[t]  
\epsfxsize=7.6cm
\begin{center}
\hspace*{-13truemm}\mbox{\epsfbox{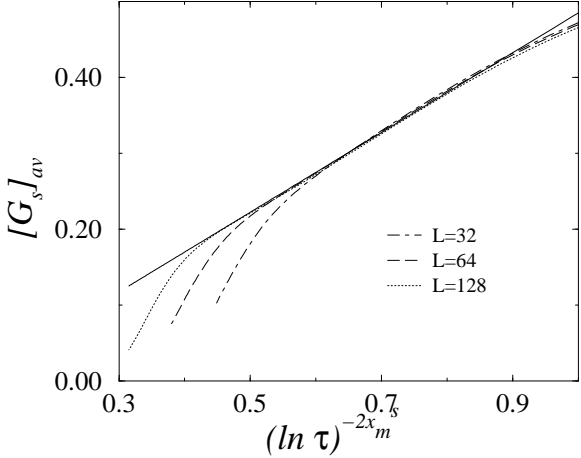}}
\end{center}
\caption{Average surface autocorrelation function for the binary distribution
in the marginal case with $A=0.5$. The solid straight line gives the expected scaling behavior in 
Eq.~(\ref{e4.15}).}
\label{fig-hvl-rand-gs-marg}
\end{figure}

The asymptotic behavior of the average surface autocorrelation function can
be  determined at criticality by scaling considerations like in a previous
work for the homogeneous case.\cite{rieger97b} Here we make use of the
fact that, as already explained for relevant perturbations, the average
autocorrelation function has the same scaling properties as the average
surface magnetization. Under a scaling transformation when lengths are
rescaled by a factor $b>1$, such that $l'=l/b$, the average surface
autocorrelation behaves as
\begin{equation}
[G_s(\ln\tau)]_{\rm{av}}=b^{-x_m^s}[G_s(\ln\tau/b^{1/2})]_{\rm{av}}\,,
\label{e4.14}
\end{equation}
where we used Eq.~(\ref{e4.6}) to relate the time and
length scales. Taking now $b=(\ln\tau)^2$ we obtain:
\begin{equation}
[G_s(\tau)]_{\rm{av}}\sim(\ln\tau)^{-2x_m^s}\sim(\ln\tau)^{-\beta_s(A)}\,.
\label{e4.15}
\end{equation}
The last expression follows from the exponent relation in Eq.~(\ref{e4.12})
where the average correlation length exponent is $\nu=2$. This scaling
behavior has also been verified by numerical calculations as shown in
Fig.~\ref{fig-hvl-rand-gs-marg}.

\subsection{Griffiths phase}

In the Griffiths phase, on the paramagnetic side of the critical
point, the dynamical properties of random quantum systems are anomalous. As a
consequence of the so-called Griffiths-McCoy singularities,\cite{griffiths69}
the different dynamical properties can be characterized by one single
parameter, the dynamical exponent $z(\delta)$, which is a continuous function
of the  quantum control parameter given in Eq.~(\ref{e1.7}) for the
homogeneous RTIM. Here we calculate $z(\delta)$ for the inhomogeneous RTIM by
considering the singular behavior of two quantities: First the probability
distribution of the gap and then the asymptotic behavior of the surface
autocorrelation function.

\begin{figure}[t] 
\epsfxsize=7.6cm
\begin{center}
\hspace*{-13truemm}\mbox{\epsfbox{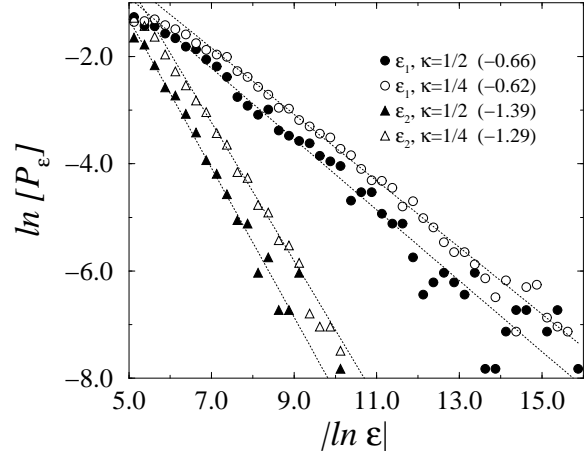}}
\end{center}
\caption{Probability distribution of the first and second gaps in the Griffiths phase
with the uniform distribution
at $h_{0}=3$ and $A=0.5$ for different values of $\kappa$. The asymptotic slopes of the
curves, obtained by least square fits, are given in parentheses.}
\label{fig-hvl-rand-z-eps1}
\end{figure}

From the distribution of the first gap one can deduce the dynamical
exponent through the asymptotic relation:\cite{igloi98a}
\begin{equation}
\lim_{\epsilon_1\to
0}{\ln\left[P(\ln\epsilon_1)\right]\over\ln\epsilon_1}={1\over z(\delta)}\,.
\label{e4.16}
\end{equation}
%
As shown in Fig.~\ref{fig-hvl-rand-z-eps1} the log-log plots of the
distributions obtained for different decay parameters $\kappa$, corresponding
to relevant and marginal perturbations, have approximately the same slopes.
Consequently, the dynamical exponent $z(\delta)$ is independent of
the inhomogeneity.  Similar results are obtained with the
probability distributions of the second gap, which is expected to obey the
same scaling relation as in Eq.~(\ref{e4.16}), however with an exponent
$z'(\delta)=z(\delta)/2$.\cite{igloi99a} As shown in Fig.~\ref{fig-hvl-rand-z-eps1}, 
the probability distribution of the second gap is also insensitive to the
presence of the inhomogeneity and the value of the exponent $z'(\delta)$,
deduced from the slope, is in good agreement with the analytical
result in Eq.~(\ref{e1.7}) with $A=0$.

Next we study the average surface autocorrelation function, the decay of
which is connected to $z(\delta)$ through Eq.~(\ref{e1.6}).
\begin{figure}[t] 
\epsfxsize=7.6cm
\begin{center}
\hspace*{-13truemm}\mbox{\epsfbox{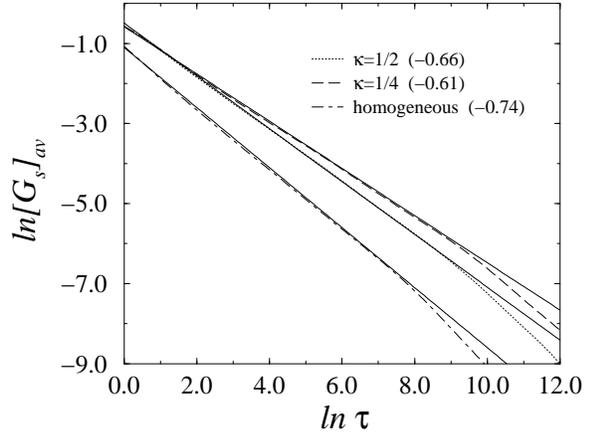}}
\end{center}
\caption{Average surface autocorrelation function in the Griffiths phase with the uniform distribution
at $h_{0}=3$, $A=0.5$ and for different values of $\kappa$. The asymptotic slopes of
the curves are given in parentheses.}
\label{fig-hvl-rand-z-gs}
\end{figure}
As shown in Fig.~\ref{fig-hvl-rand-z-gs}, the value of the decay exponent 
$z(\delta)$ has only a weak dependence
of the inhomogeneity exponent 
$\kappa$, both for relevant and marginal inhomogeneities, due to the finite size effects
which are expected to be larger when $\kappa$ is decreasing. 
Since the variation of $z$ with the inhomogeneity is
decreasing when the size of the system increases, we expect that the
surface autocorrelations are also asymptotycally independent of $\kappa$.
The value of $z(\delta)$ is consistent with that obtained from the distribution
of the gap and corresponds to the analytical bulk result in Eq.(\ref{e1.7}).

Thus we conclude that the presence of inhomogeneous disorder does not modify
the form of the Griffiths-McCoy singularities and the dynamical exponent of
the inhomogeneous RTIM is the same as for the homogeneous one.

\section{Discussion}

In this paper the influence of an inhomogeneous distribution of the
disorder in the vicinity of the surface of the RTIM has been studied. We
considered a random version of the HvL model where, on the
average, the perturbation decays as a power $l^{-\kappa}$ of the distance
$l$ from the surface. The effect of the inhomogeneity was found to be relevant
(irrelevant) for $\kappa<1/2$ ($\kappa>1/2$).

The relevance-irrelevance criterion, which was shown to follow from a simple
RW argument, can be generalized by making use of a more traditional
approach.\cite{cordery82} Near the bulk critical point, the charateristic
length scale of the unperturbed system is set by the average correlation
length $\xi\sim\delta^{-\nu}$. The strength of the perturbation at a length
scale $l$ is measured by the mean change of the control parameter,
$\overline{\delta}_A(l)=1/l\sum_{j=1}^lAl^{-\kappa}\sim Al^{-\kappa}$. This
has to be compared, at the scale of the correlation length $\xi$, to the
deviation from the critical point $\delta$. Forming the ratio
$\overline{\delta}_A(\xi)/\delta\sim A\xi^{-\kappa+1/\nu}$, we verify that
the relative strength of the perturbation diverges (vanishes) at the
critical point when $\kappa<1/\nu$ ($\kappa>1/\nu$). For the RTIM this is
equivalent to the RW result of Section III since $\nu=2$ in this
case.

Our next remark concerns {\it correlated  disorder} in the RTIM, when the
distribution of the couplings and/or transverse fields at different sites
are algebraically correlated with, for example, $\langle\pi_l(J)\pi_{l+r}(J)
\rangle\sim r^{-\omega}$. This type of correlated disorder has been
recently studied in Ref.\onlinecite{rieger00}.
The relevance-irrelevance criterion has the same form  as for inhomogeneous
disorder with $\omega$ playing the role of $\kappa$. Also the relation between
relevant time and length scales, $\ln\tau_r\sim\xi^{1-\omega}$, corresponds
to the relation in Eq.~(\ref{e4.6}) for relevant inhomogeneities. However
the form of the critical singularities for the bulk and surface magnetizations
are different in the two models.

We close this Section with a discussion of a related problem, namely, the
persistence probability, $P_{\rm per}$, for a RW in an inhomogeneous 
one-dimensional random environment which is an inhomogeneous version of 
Sinai's model.\cite{sinai82} The transition probabilities, 
$w_{l,l+1}\ne w_{l+1,l}$, are random variables
and the corresponding probability distributions are inhomogeneous near the
starting point of the walk, $l=1$, deviating from the bulk distributions by an 
amount of order $l^{-\kappa}$. We make use of an exact 
mapping\cite{igloi98b,igloi99b} between the
eigenvalue problem for the fermionic excitations of the RTIM in
Eq.~(\ref{e2.4}) and the eigenvalue problem for the master equation
of the RW, with the following correspondences: 
\begin{eqnarray}
w_{l,l+1}&\leftrightarrow&J_l^2\cr
w_{l,l-1}&\leftrightarrow&h_l^2\cr
P_{\rm per}(L)&\leftrightarrow&m_s^2(L)\cr
\lambda_{\rm min}&\leftrightarrow&-\epsilon_1^2(L)\,.
\label{e5.2}
\end{eqnarray}
Here $\lambda_{\rm min}$ is the leading eigenvalue of the Fokker-Planck 
operator, $L$ is the size of the system with free boundary conditions 
for the RTIM, whereas we take absorbing boundary conditions at both ends 
for the RW. Using the relations in Eq.~(\ref{e5.2}), one can easily translate 
the results of Section IV. The inhomogeneity is relevant (irrelevant) for the
diffusive and persistence properties of the RW for $\kappa<1/2$
($\kappa>1/2$). In the case of a relevant inhomogeneity with zero global
bias, the persistence probability is either finite, for locally
enhanced transition probabilities to the right, or it vanishes as
\begin{equation}
\left[P_{\rm per}(L)\right]_{\rm av}\sim\exp\left(-{\rm
const}\,L^{1-2\kappa}\right)\,,
\label{e5.3}
\end{equation}
for locally enhanced transition probabilities to the left.The
relation between time and length scales, $\ln t\sim L^{1-\kappa}$,
leads to:
\begin{equation}
\left[P_{\rm per}(t)\right]_{\rm{av}}\sim\exp\left[-{\rm const}\,
(\ln t)^{(1-2\kappa)/(1-\kappa)}\right]\,.
\label{e5.4}
\end{equation}
For marginal inhomogeneity the average persistence is given by
\begin{equation}
\left[P_{\rm per}(t)\right]_{\rm{av}}\sim(\ln t)^{2\theta(A)}\,,
\label{e5.5}
\end{equation}
where the continuously varying exponent $\theta(A)$ is obtained
as the solution of Eq.~(\ref{e4.9}).

\acknowledgements
We are indebted to Heiko Rieger for his help in the
numerical calculations and for interesting discussions.
This work has been supported by the French-Hungarian
cooperation program Balaton (Minist\`ere des Affaires
Etrang\`eres-O.M.F.B.), the Hungarian National Research
Fund under grants No TO23642 and M 028418 and by the Hungarian Ministery
of Education under grant No FKFP 0765/1997. The Laboratoire de Physique
des Mat\'eriaux is Unit\'e Mixte de Recherche C.N.R.S. No 7556.

\end{multicols}

\end{document}